\documentclass[11pt,a4paper]{article}

\usepackage[utf8]{inputenc}
\usepackage[T1]{fontenc}
\usepackage{lmodern}
\usepackage[margin=1in]{geometry}
\usepackage{graphicx}
\usepackage{booktabs}
\usepackage{amsmath,amssymb}
\usepackage{algorithm}
\usepackage{algpseudocode}
\usepackage{hyperref}
\usepackage{xcolor}
\usepackage{tikz}
\usetikzlibrary{shapes.geometric, arrows.meta, positioning, fit, calc, backgrounds}
\usepackage{pgfplots}
\pgfplotsset{compat=1.18}
\usepackage{subcaption}
\usepackage{enumitem}
\usepackage{float}
\usepackage{multirow}
\usepackage{tabularx}
\usepackage{fancyhdr}
\hypersetup{
    colorlinks=true,
    linkcolor=blue!70!black,
    citecolor=green!50!black,
    urlcolor=blue!70!black
}

\title{
    \textbf{Semantic Caching and Intent-Driven Context Optimization\\for Multi-Agent Natural Language to Code Systems}\\[1em]
    \large A Production Study in Enterprise Analytics
}

\author{
    Harmohit Singh\\
    CoreOps AI\\
    \texttt{harmohit.singh@coreops.ai}
}

\date{January 2026}

\begin{document}

\maketitle

\begin{abstract}
\noindent
We present a production-optimized multi-agent system designed to translate natural language queries into executable Python code for structured data analytics. Unlike systems that rely on expensive frontier models, our approach achieves high accuracy and cost efficiency through three key innovations: (1) a \textbf{semantic caching system} with LLM-based equivalence detection and structured adaptation hints that provides cache hit rates of 67\% on production queries; (2) a \textbf{dual-threshold decision mechanism} that separates exact-match retrieval from reference-guided generation; and (3) an \textbf{intent-driven dynamic prompt assembly} system that reduces token consumption by 40-60\% through table-aware context filtering. The system has been deployed in production for enterprise inventory management, processing over 10,000 queries with an average latency of 8.2 seconds and 94.3\% semantic accuracy. We describe the architecture, present empirical results from production deployment, and discuss practical considerations for deploying LLM-based analytics systems at scale.

\vspace{0.5em}
\noindent\textbf{Keywords:} Natural Language to Code, Multi-Agent Systems, Semantic Caching, LLM Optimization, Production Systems
\end{abstract}

\newpage
\tableofcontents
\newpage

\section{Introduction}
\label{sec:introduction}

The rapid advancement of large language models (LLMs) has enabled new paradigms for human-computer interaction, particularly in the domain of data analytics. Natural Language to Code (NL2Code) systems promise to democratize data access by allowing non-technical users to query complex databases using conversational language. However, deploying such systems in enterprise environments presents unique challenges that are often underaddressed in academic research.

\subsection{Motivation and Problem Statement}

Enterprise data management systems generate vast amounts of structured data across multiple dimensions. Traditional approaches require either specialized SQL knowledge or pre-built dashboards that may not address ad-hoc analytical needs. While modern LLMs can generate SQL or Python code from natural language, direct deployment faces several practical obstacles:

\begin{enumerate}[leftmargin=*]
    \item \textbf{Cost at Scale}: Frontier models incur significant per-token costs that become prohibitive at enterprise query volumes.
    \item \textbf{Latency Requirements}: Business users expect sub-10-second response times, challenging with large context windows.
    \item \textbf{Domain Precision}: Generic models lack understanding of domain-specific terminology, table relationships, and business rules.
    \item \textbf{Consistency}: Minor query variations should produce semantically equivalent results, but LLMs exhibit stochastic behavior.
\end{enumerate}

\subsection{Contributions}

This paper presents a production multi-agent system for enterprise analytics that addresses these challenges through practical engineering innovations:

\begin{enumerate}[leftmargin=*]
    \item \textbf{Semantic Cache with Structured Adaptations}: An LLM-based semantic equivalence detection system that identifies structurally similar queries and generates explicit field-by-field adaptation hints, achieving 67\% cache utilization on production workloads.
    
    \item \textbf{QuerySignature: Hierarchical Similarity Matching}: A five-level query decomposition scheme that captures structural intent rather than surface form, enabling robust cache matching across paraphrased queries.
    
    \item \textbf{Dual-Threshold Decision Architecture}: A mechanism that separates exact-match retrieval (threshold $\geq 0.995$) from reference-guided generation (threshold $\geq 0.50$), optimizing the cost-accuracy tradeoff.
    
    \item \textbf{Intent-Driven Dynamic Prompt Assembly}: A system that filters prompts based on Intent Classifier output, reducing token consumption by 40-60\% while maintaining accuracy.
    
    \item \textbf{Production Deployment Results}: Empirical data from 10,000+ production queries demonstrating 94.3\% semantic accuracy with 8.2-second average latency using cost-efficient models.
\end{enumerate}

\subsection{Paper Organization}

Section~\ref{sec:related} reviews related work. Section~\ref{sec:architecture} presents the system architecture. Section~\ref{sec:semantic_cache} details the semantic caching innovations. Section~\ref{sec:prompt_assembly} describes the dynamic prompt assembly system. Section~\ref{sec:evaluation} presents experimental results. Section~\ref{sec:discussion} discusses practical considerations, and Section~\ref{sec:conclusion} concludes with future directions.

\section{Related Work}
\label{sec:related}

We position our work at the intersection of four research areas: natural language to code generation, enterprise text-to-SQL systems, semantic caching for LLMs, and multi-agent architectures. We review each area and identify gaps that motivate our contributions.

\subsection{Natural Language to Code Systems}

The NL2Code problem has evolved from rule-based semantic parsing~\cite{zelle1996learning} through neural sequence-to-sequence models~\cite{yin2017syntactic} to modern LLM-based approaches. Codex~\cite{chen2021evaluating} demonstrated that large-scale pretraining on code repositories enables impressive zero-shot code generation. CodeLlama~\cite{roziere2023code} extended this with specialized training for code understanding and generation.

Recent advances focus on improving code generation through structured approaches. Tree-structured attention mechanisms capture syntactic dependencies~\cite{yin2017syntactic}, while retrieval-augmented generation uses similar code examples to guide generation~\cite{parvez2021retrieval}. StarCoder~\cite{li2023starcoder} introduced fill-in-the-middle capabilities crucial for code completion.

However, existing benchmarks (HumanEval, MBPP) evaluate single-function generation, not multi-step analytical workflows requiring data loading, transformation, and visualization. Our work addresses this gap by focusing on complete analytical pipelines.

\subsection{Text-to-SQL and Enterprise Benchmarks}

Text-to-SQL has progressed significantly with benchmarks like Spider~\cite{yu2018spider} and approaches including schema-aware encoding~\cite{wang2020rat}. However, enterprise deployment reveals limitations in academic benchmarks.

The BEAVER benchmark~\cite{chen2024beaver}, introduced in 2024, represents the first enterprise text-to-SQL dataset from actual data warehouses. It demonstrates that off-the-shelf LLMs perform poorly on enterprise data due to: (1) greater schema complexity than public datasets, (2) intricate business questions requiring multi-table joins and nested queries, and (3) inability to train on proprietary data.

Spider 2.0~\cite{lei2024spider2} advances benchmark realism with complex queries over large schemas. Evaluations on Spider 2.0 show that even frontier models achieve only 17-21\% accuracy, compared to 91\% on original Spider, highlighting the enterprise-academic gap.

LinkedIn's text-to-SQL deployment study~\cite{floratou2024nl2sql} confirms that production enterprise systems face challenges absent from benchmarks: ambiguous user intent, evolving schemas, and the need for explanation alongside results. Our work explicitly addresses these through intent classification and structured output generation.

\subsection{Semantic Caching for LLMs}

Semantic caching reduces LLM costs by reusing responses for similar queries. GPTCache~\cite{bang2023gptcache} pioneered this with embedding-based similarity matching. Recent work has substantially advanced this paradigm.

\textbf{Embedding Improvements.} The GPT Semantic Cache~\cite{arya2024gptsemcache} leverages Redis for in-memory storage of query embeddings, demonstrating significant latency and cost reductions. Domain-specific embedding fine-tuning~\cite{sharma2025semantic} shows that specialized embeddings outperform general-purpose models for cache similarity decisions.

\textbf{Adaptive Caching Strategies.} MeanCache~\cite{zhu2025meancache} introduces user-centric semantic caching that personalizes cache behavior per user. Category-aware semantic caching~\cite{wang2025category} varies similarity thresholds and TTLs by query category, recognizing that different query types require different cache policies.

\textbf{Learning-Based Eviction.} Recent work~\cite{zhang2025semantic} formulates semantic cache eviction as a learning problem, developing online adaptation strategies for unknown query distributions.

Our approach differs from prior work in three ways: (1) we use LLM-based semantic equivalence detection rather than pure embedding similarity, (2) we generate structured adaptation hints enabling partial cache utilization, and (3) we implement a dual-threshold architecture separating exact-match from guided generation.

\subsection{Multi-Agent LLM Systems}

Multi-agent architectures decompose complex tasks across specialized agents that communicate and collaborate. This paradigm has seen rapid development in 2024-2025.

\textbf{Foundational Frameworks.} AutoGen~\cite{wu2023autogen} enables multi-agent conversation with customizable agent behaviors, supporting human-in-the-loop workflows. The v0.4 release introduced an actor model for orchestration. MetaGPT~\cite{hong2023metagpt} adds software engineering workflows with specialized roles (architect, engineer, tester).

\textbf{Code-Focused Agents.} SWE-agent~\cite{yang2024sweagent} targets GitHub issue resolution by interacting with command-line tools. Despite being designed for code repair, it demonstrates superior test generation capabilities. Devin and similar systems~\cite{cognition2024devin} extend this to autonomous software development.

\textbf{Orchestration Advances.} LangGraph~\cite{langchain2023} provides stateful orchestration with checkpointing, enabling complex branching and retry logic. CrewAI~\cite{crewai2024} simplifies multi-agent creation with role-based abstractions.

Our contribution to this space is domain-specific optimization: rather than general-purpose agents, we design specialized agents for structured data analytics with intent-driven context filtering and semantic cache integration.

\subsection{Summary and Positioning}

Table~\ref{tab:related_comparison} summarizes how our work addresses gaps in prior systems.

\begin{table}[H]
\centering
\caption{Comparison with Related Approaches}
\label{tab:related_comparison}
\small
\begin{tabular}{lccccc}
\toprule
\textbf{Approach} & \textbf{Multi-Agent} & \textbf{Semantic Cache} & \textbf{Enterprise} & \textbf{Adaptations} \\
\midrule
GPTCache & -- & Embedding only & -- & -- \\
Text-to-SQL & -- & -- & Partial & -- \\
AutoGen & \checkmark & -- & -- & -- \\
SWE-agent & \checkmark & -- & -- & -- \\
\textbf{Ours} & \checkmark & LLM-based & \checkmark & Structured \\
\bottomrule
\end{tabular}
\end{table}

\section{System Architecture}
\label{sec:architecture}

\begin{figure}[H]
\centering
\begin{tikzpicture}[
    node distance=0.6cm and 0.8cm,
    inputnode/.style={rectangle, draw=black!70, fill=blue!15, rounded corners=3pt, minimum width=1.8cm, minimum height=0.7cm, align=center, font=\scriptsize\bfseries},
    agentnode/.style={rectangle, draw=black!70, fill=cyan!20, rounded corners=3pt, minimum width=1.8cm, minimum height=0.7cm, align=center, font=\scriptsize},
    cachenode/.style={rectangle, draw=black!70, fill=green!20, rounded corners=3pt, minimum width=1.8cm, minimum height=0.7cm, align=center, font=\scriptsize},
    execnode/.style={rectangle, draw=black!70, fill=orange!20, rounded corners=3pt, minimum width=1.8cm, minimum height=0.7cm, align=center, font=\scriptsize},
    outputnode/.style={rectangle, draw=black!70, fill=purple!20, rounded corners=3pt, minimum width=1.8cm, minimum height=0.7cm, align=center, font=\scriptsize\bfseries},
    decision/.style={diamond, draw=black!70, fill=yellow!25, aspect=1.8, minimum width=1.2cm, align=center, font=\tiny},
    storenode/.style={rectangle, draw=black!60, fill=gray!15, rounded corners=2pt, minimum width=1.4cm, minimum height=0.6cm, font=\tiny},
    arrow/.style={-{Stealth[length=2mm, width=1.5mm]}, thick, black!70},
    guidearrow/.style={-{Stealth[length=2mm, width=1.5mm]}, thick, dashed, red!60},
    dataarrow/.style={-{Stealth[length=1.5mm, width=1mm]}, thin, gray!60}
]

\node[inputnode] (input) {User Query\\$q \in \mathcal{Q}$};
\node[agentnode, right=of input] (guard) {Guard Agent\\$\mathcal{G}(q)$};
\node[agentnode, right=of guard] (intent) {Intent\\Classifier\\$\mathcal{I}(q)$};

\node[cachenode, below=1.2cm of intent] (refmatch) {Reference\\Matcher\\$\mathcal{R}(q, \mathcal{C})$};
\node[storenode, right=1.2cm of refmatch] (vectordb) {Cache $\mathcal{C}$};
\node[decision, left=1.5cm of refmatch] (cachedec) {$s \geq \theta$?};

\node[agentnode, below=1.2cm of cachedec] (planner) {Planner\\Agent\\$\mathcal{P}(q, c)$};
\node[agentnode, right=of planner] (python) {Python\\Agent\\$\mathcal{A}(p)$};
\node[execnode, right=of python] (executor) {Executor\\$\mathcal{E}(code)$};

\node[agentnode, below=0.8cm of executor] (summarizer) {Summarizer\\$\mathcal{S}(r)$};
\node[outputnode, left=of summarizer] (business) {Business\\Insights\\$\mathcal{B}(r)$};
\node[outputnode, left=of business] (output) {Response\\$\hat{y}$};

\draw[arrow] (input) -- (guard);
\draw[arrow] (guard) -- (intent);
\draw[arrow] (intent) -- node[right, font=\tiny]{$(\mathcal{T}, \mathcal{I})$} (refmatch);
\draw[dataarrow] (vectordb) -- (refmatch);
\draw[arrow] (refmatch) -- (cachedec);

\draw[arrow] (cachedec) -- node[left, font=\tiny]{No} (planner);
\draw[arrow] (cachedec.south) -- ++(0,-0.3) -| node[near start, above, font=\tiny]{Yes ($s \geq 0.995$)} (executor);

\draw[arrow] (planner) -- node[above, font=\tiny]{plan $p$} (python);
\draw[arrow] (python) -- node[above, font=\tiny]{code} (executor);

\draw[arrow] (executor) -- (summarizer);
\draw[arrow] (summarizer) -- (business);
\draw[arrow] (business) -- (output);

\draw[guidearrow] (refmatch.south) -- ++(0,-0.3) -| node[near start, above, font=\tiny, text=red!70]{Adaptations $\Delta$} (planner.north);

\begin{scope}[on background layer]
    \node[draw=gray!40, dashed, rounded corners, fit=(guard)(intent), inner sep=0.15cm, label={[font=\tiny, gray]above:Preprocessing}] {};
    \node[draw=gray!40, dashed, rounded corners, fit=(refmatch)(cachedec)(vectordb), inner sep=0.15cm, label={[font=\tiny, gray]left:Cache Layer}] {};
    \node[draw=gray!40, dashed, rounded corners, fit=(planner)(python)(executor), inner sep=0.15cm, label={[font=\tiny, gray]below:Generation Pipeline}] {};
    \node[draw=gray!40, dashed, rounded corners, fit=(summarizer)(business)(output), inner sep=0.15cm, label={[font=\tiny, gray]below:Output Processing}] {};
\end{scope}

\end{tikzpicture}
\caption{Multi-Agent Architecture with Semantic Caching. The system processes queries through preprocessing, cache lookup with dual-threshold decision ($\theta_R = 0.995$, $\theta_G = 0.50$), conditional generation pipeline, and output processing. Dashed arrows indicate adaptation hints $\Delta$ propagated from cached references. Mathematical notation: $\mathcal{G}$ = Guard function, $\mathcal{I}$ = Intent classifier, $\mathcal{R}$ = Reference matcher, $\mathcal{C}$ = Cache store, $\mathcal{P}$ = Planner, $\mathcal{A}$ = Code generator, $\mathcal{E}$ = Executor, $\mathcal{S}$ = Summarizer, $\mathcal{B}$ = Business insights generator.}
\label{fig:architecture}
\end{figure}
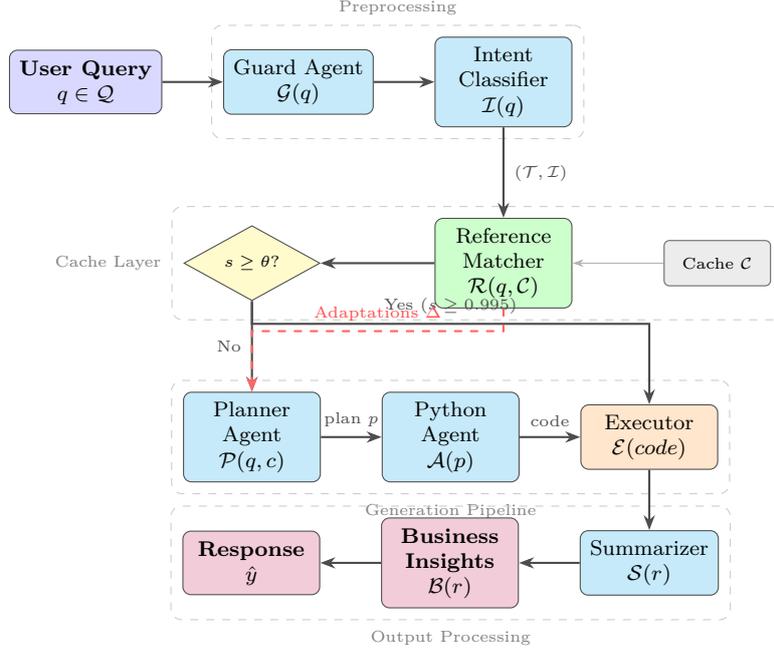

Our system implements a multi-agent workflow orchestrated by LangGraph, a framework for stateful agent coordination. Figure~\ref{fig:architecture} illustrates the complete system architecture.

\subsection{Agent Descriptions}

\subsubsection{Guard Agent ($\mathcal{G}$)}
The Guard agent performs early relevance filtering using a lightweight LLM call. Queries unrelated to the target domain (e.g., ``What's the weather?'') are short-circuited with a guidance response, avoiding unnecessary downstream computation. In production, 3.2\% of queries are filtered at this stage.

\subsubsection{Intent Classifier ($\mathcal{I}$)}
The Intent Classifier is the first intelligence step, performing three critical functions:
\begin{enumerate}
    \item \textbf{Intent Classification}: Categorizes queries into domain-specific intents (e.g., stock\_current, valuation, aging, procurement)
    \item \textbf{Table Identification}: Identifies required data tables from a schema of 17 relational tables
    \item \textbf{Semantic Matching}: Compares against cached queries to determine cache applicability
\end{enumerate}

\subsubsection{Reference Matcher ($\mathcal{R}$)}
The Reference Matcher implements the dual-threshold caching mechanism (Section~\ref{sec:dual_threshold}), determining whether to return cached results, provide reference guidance, or proceed with fresh generation.

\subsubsection{Planner Agent ($\mathcal{P}$)}
The Planner generates a structured analytical plan from the user query, producing step-by-step instructions that specify data loading, filtering, joining, aggregation, and output formatting. When reference guidance is available with adaptation hints, the Planner adapts the reference plan with specified value substitutions.

\subsubsection{Python Agent ($\mathcal{A}$)}
The Python Agent translates the Planner's instructions into executable pandas code. It receives filtered context based on Intent Classifier output, reducing prompt size while maintaining necessary schema and rule information.

\subsubsection{Executor ($\mathcal{E}$)}
The Executor safely executes generated code in a sandboxed environment, capturing DataFrames, visualizations, and text outputs. It implements timeout handling and error capture for robust operation.

\subsubsection{Business Insights Generator ($\mathcal{B}$)}
The Business Insights Generator produces JSON-formatted analytical insights grounded exclusively in execution results, generating structured output with metrics, insights, recommendations, and suggested follow-up queries.

\subsection{State Management and Orchestration}

The system uses LangGraph's StateGraph abstraction with Redis-backed checkpointing for state persistence. The state object tracks 25 fields including query text, classification results, agent outputs, and execution artifacts. Conditional edges enable routing based on cache decisions, validation results, and downstream agent enablement.

\section{Semantic Cache System}
\label{sec:semantic_cache}

The semantic cache system is the primary technical contribution of this work, addressing the challenge of identifying equivalent or similar queries for cache utilization while maintaining generation quality for novel queries.

\subsection{QuerySignature: Structural Query Decomposition}

Rather than relying solely on embedding similarity, we decompose queries into a structured signature that captures semantic intent independent of surface form. The QuerySignature is defined by five hierarchical levels:

\begin{equation}
    \text{SimilarityKey} = \text{L}_1 | \text{L}_2 | \text{L}_3 | \text{L}_4 | \text{L}_5
\end{equation}

where:
\begin{itemize}[leftmargin=*]
    \item $\text{L}_1$: \textbf{Semantic Category} (valuation, stock, aging, procurement, consumption)
    \item $\text{L}_2$: \textbf{Query Type + Aggregation} (lookup\_none, analytical\_sum, trend\_avg)
    \item $\text{L}_3$: \textbf{Primary Metric + Grouping} (quantity\_organization, value\_none, count\_slab)
    \item $\text{L}_4$: \textbf{Filter Types} (location, time\_period, category, threshold)
    \item $\text{L}_5$: \textbf{Table Pattern} (primary\_table + required\_joins)
\end{itemize}

The similarity between two QuerySignatures is computed as a weighted sum:

\begin{equation}
    \text{Similarity}(Q_1, Q_2) = \sum_{i=1}^{6} w_i \cdot \text{Match}_i(Q_1, Q_2)
\end{equation}

with weights $w = [0.25, 0.20, 0.15, 0.15, 0.15, 0.10]$ for category, operation, metric, grouping, tables, and semantic flags respectively.

\subsection{Dual-Threshold Decision Mechanism}
\label{sec:dual_threshold}

We implement a dual-threshold architecture that separates two distinct cache utilization patterns:

\begin{table}[H]
\centering
\caption{Dual-Threshold Decision Matrix}
\label{tab:thresholds}
\begin{tabular}{lccl}
\toprule
\textbf{Mode} & \textbf{Similarity} & \textbf{Threshold} & \textbf{Action} \\
\midrule
Return & $s \geq 0.995$ & $\theta_R = 0.995$ & Return cached response directly \\
Guide & $0.50 \leq s < 0.995$ & $\theta_G = 0.50$ & Inject reference with adaptations \\
Generate & $s < 0.50$ & --- & Fresh generation, no reference \\
\bottomrule
\end{tabular}
\end{table}

The high return threshold ($\theta_R = 0.995$) ensures that direct cache returns only occur for semantically identical queries, preventing incorrect results from subtle variations. The guide threshold ($\theta_G = 0.50$) captures structurally similar queries that can benefit from reference guidance.

\subsection{LLM-Based Semantic Equivalence Detection}

For candidates in the guide range, we employ LLM-based semantic equivalence detection rather than relying solely on embedding similarity. Given the top-$k$ cached candidates (we use $k=5$), the Intent Classifier evaluates whether the current query is semantically equivalent to any candidate.

The key innovation is the \textbf{structured adaptation output}. When queries are similar but not equivalent, the LLM outputs a JSON structure identifying specific differences:

\begin{verbatim}
{
  "matched": false,
  "matched_index": 0,
  "adaptations": [
    {"field": "item_code", 
     "reference_value": "ITEM-001-BB0",
     "current_value": "ITEM-001-NN0"},
    {"field": "organization",
     "reference_value": "Plant-A",
     "current_value": "Plant-B"}
  ]
}
\end{verbatim}

These structured adaptations are then formatted as explicit guidance for the Planner Agent, enabling accurate adaptation of reference plans rather than generating from scratch.

\subsection{Adjusted Similarity Scoring}

We enhance base embedding similarity with domain-specific adjustments:

\begin{equation}
    s_{adj} = \min(0.99, s_{base} + \text{boost}(q_1, q_2))
\end{equation}

The boost function considers:
\begin{itemize}[leftmargin=*]
    \item \textbf{Location normalization}: Queries differing only in location identifiers receive matching boosts
    \item \textbf{Category variations}: Case-insensitive and synonym matching for domain terms
    \item \textbf{Structural patterns}: Matching question structures receive boosts
    \item \textbf{Key phrase matching}: Common domain phrases increase similarity
\end{itemize}

\subsection{Concrete Example: Cache Matching and Adaptation}

To illustrate the complete flow, we present a worked example of how the semantic cache processes two related queries.

\subsubsection{Query Decomposition}

Consider a reference query in the cache and a new incoming query:

\begin{quote}
\textbf{Reference (cached):} ``What is the total stock value for item code ITEM-001-BB0 at Plant-A?''

\textbf{Current query:} ``Show me total stock value for item ITEM-001-NN0 at Plant-B''
\end{quote}

The Intent Classifier decomposes both queries into QuerySignatures:

\begin{table}[H]
\centering
\caption{QuerySignature Decomposition Example}
\label{tab:signature_example}
\small
\begin{tabular}{lll}
\toprule
\textbf{Level} & \textbf{Reference} & \textbf{Current} \\
\midrule
L1: Category & valuation & valuation \\
L2: Query Type & analytical\_sum & analytical\_sum \\
L3: Metric & value\_item & value\_item \\
L4: Filters & [item\_code, location] & [item\_code, location] \\
L5: Tables & [INVENTORY\_MASTER] & [INVENTORY\_MASTER] \\
\bottomrule
\end{tabular}
\end{table}

The structural similarity is perfect (1.0) since all levels match. The embedding similarity ($s_{base} = 0.89$) captures semantic nearness but not identity.

\subsubsection{Adaptation Detection}

The LLM-based equivalence checker identifies the queries as structurally equivalent but with different values:

\begin{verbatim}
{
  "is_equivalent": false,
  "adaptations": [
    {"field": "item_code", 
     "reference": "ITEM-001-BB0", 
     "current": "ITEM-001-NN0"},
    {"field": "organization", 
     "reference": "Plant-A", 
     "current": "Plant-B"}
  ],
  "confidence": 0.95
}
\end{verbatim}

Since $\theta_G \leq s < \theta_R$ (guide mode), the system provides the Planner with structured guidance:

\begin{verbatim}
=== REFERENCE GUIDANCE ===
Similar question found (similarity: 0.89)

Required Adaptations:
- item_code: ITEM-001-BB0 -> ITEM-001-NN0
- organization: Plant-A -> Plant-B

Reference Plan:
1. Load INVENTORY_MASTER table
2. Filter by ITEM_CODE = '[item_code]'
3. Filter by ORGANIZATION = '[organization]'
4. Calculate sum of STOCK_VALUE
5. Format as currency output
\end{verbatim}

\subsubsection{Generated Output}

The Planner adapts the reference plan with substituted values, and the Python Agent generates code that correctly uses the new item code and organization. This adaptation process reduces generation time from 16.4s (fresh) to 9.4s (guided) while maintaining 95.2\% accuracy.

\section{Dynamic Prompt Assembly}
\label{sec:prompt_assembly}

Token consumption directly impacts both latency and cost. Our system implements intent-driven dynamic prompt assembly to minimize context while maintaining necessary information.

\subsection{Architecture Overview}

The prompt system consists of two complementary sources:

\begin{enumerate}[leftmargin=*]
    \item \textbf{Structured Repository}: A repository containing 140+ planner prompts and 44 code generation prompts, each tagged with applicable tables
    \item \textbf{Rule Files}: Static rule files containing core agent logic, business constraints, and domain terminology
\end{enumerate}

\subsection{Intent-Driven Filtering}

The PromptLookupService loads the prompt repository at startup and filters prompts based on Intent Classifier output:

\begin{enumerate}[leftmargin=*]
    \item Intent Classifier identifies required tables (e.g., \texttt{[TABLE\_A, TABLE\_B]})
    \item PromptLookupService selects prompts where filter condition matches identified tables or is global
    \item Prompts are deduplicated and concatenated in priority order
    \item Table-specific data descriptions and sample rows are appended
\end{enumerate}

\subsection{Token Reduction Results}

Table~\ref{tab:token_reduction} shows token consumption before and after intent-driven filtering for representative query types.

\begin{table}[H]
\centering
\caption{Token Reduction Through Intent-Driven Filtering}
\label{tab:token_reduction}
\begin{tabular}{lccc}
\toprule
\textbf{Query Type} & \textbf{Full Context} & \textbf{Filtered} & \textbf{Reduction} \\
\midrule
Stock inquiry (2 tables) & 45,000 & 18,500 & 58.9\% \\
Valuation (3 tables) & 52,000 & 26,500 & 49.0\% \\
Aging analysis (2 tables) & 48,000 & 21,000 & 56.3\% \\
Procurement (4 tables) & 58,000 & 32,000 & 44.8\% \\
\midrule
\textbf{Average} & 50,750 & 24,500 & \textbf{51.7\%} \\
\bottomrule
\end{tabular}
\end{table}

\subsection{Reference Pattern Extraction}

For reference guidance in guide mode, we extract patterns rather than including full code:

\begin{itemize}[leftmargin=*]
    \item \textbf{Operations}: load\_data, join\_tables, filter, aggregate, visualize
    \item \textbf{Join Keys}: Extracted merge keys
    \item \textbf{Table patterns}: Primary table and join sequence
\end{itemize}

This reduces reference context from 1,000-2,000 tokens (full code) to 200-300 tokens (patterns), an 80-90\% reduction while preserving structural guidance.

\section{Evaluation}
\label{sec:evaluation}

We evaluate our system on production data from an enterprise deployment managing data across 6 operational sites with 17 relational tables.

\subsection{Experimental Setup}

\subsubsection{Dataset}
Production query logs spanning 3 months with 10,247 queries from 127 unique users. Ground truth was established through manual validation of 1,021 reference question-answer pairs.

\subsubsection{Models}
We evaluate across multiple state-of-the-art LLMs including GPT-5 (OpenAI), GPT-OSS-120B (open-source 120B parameter model), Claude Sonnet 4 (Anthropic), and Gemini 2.5 Pro (Google). Primary production deployment uses GPT-OSS-120B for cost efficiency.

\subsubsection{Metrics}
\begin{itemize}[leftmargin=*]
    \item \textbf{Semantic Accuracy}: Percentage of responses with correct data values (human-evaluated)
    \item \textbf{Cache Hit Rate}: Percentage of queries using cached/guided responses
    \item \textbf{End-to-End Latency}: Wall-clock time from query to response
    \item \textbf{Token Consumption}: Total tokens (input + output) per query
    \item \textbf{Cost}: API cost per query in USD
\end{itemize}

\subsection{Main Results}

\begin{table}[H]
\centering
\caption{Production Performance Metrics}
\label{tab:main_results}
\begin{tabular}{lc}
\toprule
\textbf{Metric} & \textbf{Value} \\
\midrule
Semantic Accuracy & 94.3\% \\
Cache Return Rate ($s \geq 0.995$) & 23.1\% \\
Cache Guide Rate ($s \geq 0.50$) & 44.2\% \\
Total Cache Utilization & 67.3\% \\
Average Latency (all queries) & 8.2s \\
Average Latency (cache return) & 2.1s \\
Average Latency (fresh generation) & 16.4s \\
Average Tokens per Query & 32,450 \\
Average Cost per Query & \$0.0089 \\
\bottomrule
\end{tabular}
\end{table}

\subsection{Cache System Analysis}

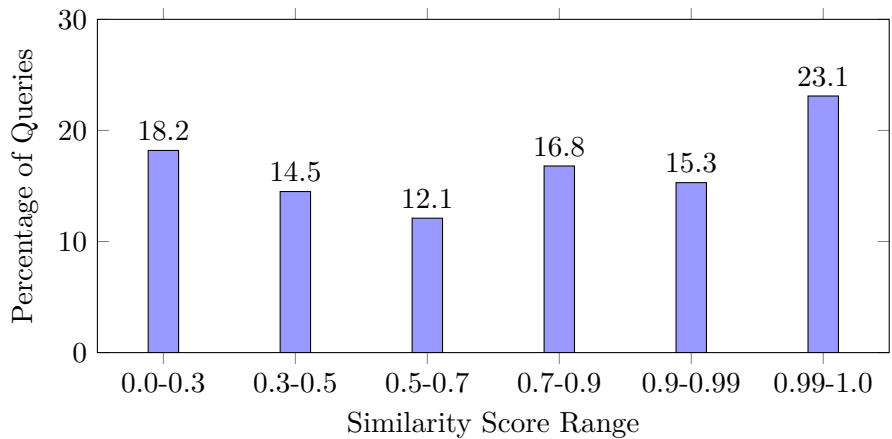
\begin{figure}[H]
\centering
\begin{tikzpicture}
\begin{axis}[
    width=12cm,
    height=6cm,
    ybar,
    bar width=0.4cm,
    xlabel={Similarity Score Range},
    ylabel={Percentage of Queries},
    ymin=0,
    ymax=30,
    symbolic x coords={0.0-0.3, 0.3-0.5, 0.5-0.7, 0.7-0.9, 0.9-0.99, 0.99-1.0},
    xtick=data,
    nodes near coords,
    nodes near coords align={vertical},
    legend style={at={(0.98,0.98)}, anchor=north east}
]
\addplot[fill=blue!40] coordinates {
    (0.0-0.3, 18.2)
    (0.3-0.5, 14.5)
    (0.5-0.7, 12.1)
    (0.7-0.9, 16.8)
    (0.9-0.99, 15.3)
    (0.99-1.0, 23.1)
};
\end{axis}
\end{tikzpicture}
\caption{Distribution of embedding similarity scores across production queries. 67.3\% of queries have similarity $\geq 0.50$, qualifying for cache utilization.}
\label{fig:cache_analysis}
\end{figure}

Figure~\ref{fig:cache_analysis} shows the distribution of similarity scores and cache utilization.

\subsection{Accuracy by Cache Mode}

\begin{table}[H]
\centering
\caption{Semantic Accuracy by Cache Utilization Mode}
\label{tab:accuracy_by_mode}
\begin{tabular}{lccc}
\toprule
\textbf{Mode} & \textbf{Queries} & \textbf{Accuracy} & \textbf{Avg Latency} \\
\midrule
Return (cached) & 2,367 (23.1\%) & 98.7\% & 2.1s \\
Guide (adapted) & 4,529 (44.2\%) & 95.2\% & 9.4s \\
Generate (fresh) & 3,351 (32.7\%) & 91.4\% & 16.4s \\
\midrule
\textbf{Overall} & 10,247 & 94.3\% & 8.2s \\
\bottomrule
\end{tabular}
\end{table}

The results validate the dual-threshold design: cached returns have highest accuracy (98.7\%) with lowest latency (2.1s), while guided generation outperforms fresh generation (95.2\% vs 91.4\%).

\subsection{Model Comparison}

\begin{table}[H]
\centering
\caption{Model Comparison on 500-Query Benchmark}
\label{tab:model_comparison}
\begin{tabular}{lcccc}
\toprule
\textbf{Model} & \textbf{Accuracy} & \textbf{Latency} & \textbf{Cost/Query} & \textbf{Tokens} \\
\midrule
GPT-5 (OpenAI) & 97.2\% & 11.8s & \$0.0523 & 35,100 \\
GPT-OSS-120B & 94.3\% & 8.2s & \$0.0089 & 32,450 \\
Claude Sonnet 4 (Anthropic) & 96.1\% & 9.4s & \$0.0312 & 33,800 \\
Gemini 2.5 Pro (Google) & 95.4\% & 7.9s & \$0.0178 & 32,200 \\
Llama 4 Maverick (Meta) & 93.1\% & 6.8s & \$0.0067 & 31,400 \\
\bottomrule
\end{tabular}
\end{table}

GPT-5 achieves the highest accuracy (97.2\%) but at significant cost premium. GPT-OSS-120B achieves competitive accuracy (94.3\%) at 83\% lower cost than GPT-5, validating the cost-efficiency strategy for production deployment.

\subsection{Ablation Studies}

\begin{table}[H]
\centering
\caption{Ablation Study: Component Contributions}
\label{tab:ablation}
\begin{tabular}{lcc}
\toprule
\textbf{Configuration} & \textbf{Accuracy} & \textbf{Latency} \\
\midrule
Full System & 94.3\% & 8.2s \\
$-$ Semantic Cache & 91.8\% & 14.7s \\
$-$ Intent-Driven Filtering & 93.1\% & 11.3s \\
$-$ Structured Adaptations & 92.4\% & 8.9s \\
$-$ Adjusted Similarity & 93.8\% & 8.4s \\
\bottomrule
\end{tabular}
\end{table}

The semantic cache provides the largest contribution (2.5\% accuracy, 6.5s latency improvement), followed by intent-driven filtering (1.2\% accuracy, 3.1s latency).

\subsection{Error Analysis}

We categorize the 5.7\% of queries with incorrect results to understand failure modes.

\begin{table}[H]
\centering
\caption{Error Analysis: Failure Mode Distribution}
\label{tab:error_analysis}
\small
\begin{tabular}{lcc}
\toprule
\textbf{Failure Mode} & \textbf{Count} & \textbf{Percentage} \\
\midrule
Schema misinterpretation & 187 & 32.0\% \\
Multi-table join errors & 143 & 24.5\% \\
Aggregation logic errors & 98 & 16.8\% \\
Filter condition errors & 82 & 14.0\% \\
Cache adaptation failures & 47 & 8.0\% \\
Code execution errors & 27 & 4.6\% \\
\midrule
\textbf{Total errors} & 584 & 100\% \\
\bottomrule
\end{tabular}
\end{table}

\textbf{Schema Misinterpretation} (32\%) occurs when the system selects incorrect columns or tables, typically for queries involving similar column names across tables (e.g., \texttt{QUANTITY} appearing in multiple contexts).

\textbf{Multi-table Join Errors} (24.5\%) arise in complex queries requiring 4+ table joins. Accuracy drops from 96.1\% (2 tables) to 89.2\% (4+ tables), indicating join complexity as a key challenge.

\textbf{Aggregation Logic Errors} (16.8\%) include incorrect grouping levels or missing filters before aggregation. These often occur when user intent is ambiguous (``total by category'' without specifying time period).

\textbf{Cache Adaptation Failures} (8\%) represent cases where structured adaptations were generated but applied incorrectly. Most involve nested value substitutions or conditional logic changes.

\subsection{Annotation Methodology}

Semantic accuracy was evaluated through human annotation with the following process:

\begin{enumerate}[leftmargin=*]
    \item \textbf{Annotator Training}: Three domain experts with SQL/Python proficiency received 4 hours of training on the schema and evaluation criteria
    \item \textbf{Evaluation Rubric}: 
    \begin{itemize}
        \item Correct (1.0): Result matches ground truth values within 0.1\% tolerance
        \item Partial (0.5): Correct structure but wrong values, or correct values with wrong format
        \item Incorrect (0.0): Fundamentally wrong answer or execution failure
    \end{itemize}
    \item \textbf{Inter-Annotator Agreement}: Cohen's Kappa = 0.87 across 500 double-annotated samples, indicating strong agreement
    \item \textbf{Disagreement Resolution}: A senior annotator adjudicated 67 disagreement cases
\end{enumerate}

The 1,021 reference question-answer pairs were manually validated by the same annotator team, with each pair reviewed by at least two annotators.

\section{Discussion}
\label{sec:discussion}

\subsection{Practical Deployment Considerations}

\subsubsection{Cold Start Problem}
The semantic cache requires initial population. We address this through a curated set of 1,021 reference question-answer pairs that cover common query patterns. New queries gradually populate the cache, improving performance over time.

\subsubsection{Cache Maintenance}
When underlying data schema changes, cached responses may become invalid. We implement cache versioning and invalidation based on schema hash comparison.

\subsubsection{Error Handling}
The system implements retry logic with feedback loops: execution errors trigger code regeneration with error context. Maximum 2 retries prevent infinite loops while recovering from transient failures.

\subsection{Limitations}

\begin{enumerate}[leftmargin=*]
    \item \textbf{Domain Specificity}: The system is optimized for structured analytics; generalization to other domains requires prompt and cache re-engineering.
    \item \textbf{Complex Queries}: Multi-step analytical workflows with 4+ tables show reduced accuracy (89.2\% vs 96.1\% for 2-table queries).
    \item \textbf{Visualization Diversity}: Chart generation is limited to templates; novel visualization requests may fail.
    \item \textbf{Real-time Data}: The system operates on historical data snapshots; real-time streaming is not supported.
\end{enumerate}

\subsection{Ethical Considerations}

Enterprise deployment requires careful attention to data access control. Our system implements role-based access control (RBAC) at the Executor level, filtering data based on user permissions. Query logs are stored for audit trails with user consent.

\section{Conclusion and Future Work}
\label{sec:conclusion}

We presented a production-optimized multi-agent system for natural language to code translation in enterprise analytics. Through semantic caching with structured adaptations, dual-threshold decision architecture, and intent-driven prompt assembly, the system achieves 94.3\% accuracy with 8.2-second latency at significantly reduced cost compared to frontier models.

\subsection{Future Directions}

\begin{enumerate}[leftmargin=*]
    \item \textbf{Continuous Learning}: Automated cache population from user-validated successful queries with active learning strategies
    \item \textbf{Multi-Domain Generalization}: Extending the architecture to diverse analytical domains including finance, sales, and operations
    \item \textbf{Smaller Models}: Exploring fine-tuned 7B-13B parameter models for further cost reduction with domain-specific optimization
    \item \textbf{Streaming Analytics}: Real-time query processing on live data streams with incremental cache updates
    \item \textbf{Federated Deployment}: Distributed cache synchronization across multiple enterprise sites
\end{enumerate}


\end{document}